\documentclass{aa}                        % default manuscript style (AA)
\usepackage{graphicx, latexsym}      % Prefered graphics (AA)
\usepackage{natbib}                  % Bibliography package
\bibpunct{(}{)}{;}{a}{}{,}           % to follow the A&A style

\begin{document}

\title{Axi-symmetric Models of B[e] Supergiants:  \\ I. The Effective Temperature 
and Mass-loss Dependence of the Hydrogen and Helium Ionization Structure}

\author{J. Zsarg\'{o}\inst{1}
\and D. J. Hillier\inst{1}
\and L. N. Georgiev\inst{2}}

\institute{ Dept. of Physics and Astronomy,
University of Pittsburgh, 3941 O'Hara St.,
Pittsburgh, PA 15260, USA
\and Instituto de Astronomia,
Universidad Nacional Autonoma de Mexico (UNAM),
CD. Universitaria, Apartado Postal 70-264,
04510, M\'{e}xico DF, M\'{e}xico}

\offprints{J.Zsarg\'{o}, \email{jzsargo@astro.phyast.pitt.edu}}

\date{Received DD MM 200Y/ Accepted DD MM 200Y}

\abstract
{}
{We calculate the hydrogen and helium ionization in B[e] envelopes and explore
their dependence on mass-loss and effective temperature. 
We also present simulated observations of the H$\alpha$ emission line and
the \ion{C}{IV} $\lambda\lambda$1550 doublet, and study their behavior.
This paper reports our first results in an ongoing study of B[e] supergiants,
and provides a glimpse on the ionization of the most important elements in 
self-consistent numerical simulations.
}
{Our newly developed 2D stellar atmosphere code, ASTAROTH, was used for
the numerical simulations. The code self-consistently solves for the continuum radiation,
non-LTE level populations, and electron temperature in axi-symmetric stellar envelopes.
Observed profiles were calculated by an auxiliary program developed separately from
ASTAROTH.
}
{In all but one of our models, H remained fully ionized ---  only  for \.{M}$>$10$^{-5}$~M$_{\odot}$~yr$^{-1}$
and T$_{eff} \leq$18,000~K did we obtain a neutral H disk, and then only for radii beyond $3R_*$.
Due to ionizations from excited states it is much more difficult to get a H neutral disk than
indicated by previous analytical calculations. 
Near the poles, the ionization is high in all models, while helium recombined in the equatorial regions for all
but our lowest mass-loss rate (10$^{-6}$~M$_{\odot}$~yr$^{-1}$). 
Although the model parameters were not adjusted to provide fits to any particular star,
the theoretical profiles show some features seen in the profiles of R126. These include the partially
resolved double peaked profile of H$\alpha$, and the weak emission associated with the UV C\,{\sc iv}
resonance line.
}
{}
\keywords{Physical data and processes:  Radiative transfer -- Stars: early-type --
                   Stars: atmospheres -- Stars: mass-loss}

\titlerunning{Ionization in the wind of sgB[e] stars}
\authorrunning{Zsarg\'{o} et al.}

\maketitle

\section{Introduction}\label{section:intro}

Massive stars are very important constituents of the Universe despite their 
low share of galactic masses.
They are the primary sources of elements heavier than Li and power the internal
evolution of galaxies by their radiation, winds, and explosions.
Many of these stars, and their immediate environment, cannot be fully understood 
in the context of plane-parallel or spherical models, and their multidimensional
nature has to be taken into account if their evolution or physical state is
to be adequately described.
The origin of their asphericity is either fast rotation, the presence 
of a dynamically important magnetic field, or their interaction with a companion.

Three groups of such stars are particularly relevant for this paper.
These are the classical Be stars, the Luminous Blue Variables (LBV), and the B[e] 
supergiants (sgB[e]), all of which share many similar characteristics.
The 2D or 3D nature of these objects was recognized through both observations 
and theoretical calculations.
For example, the presence of disks around classical Be stars was inferred from line 
modeling and polarimetric studies \citep{poe78a, poe78b}, and has been confirmed 
by interferometric observations \citep{ste95, qui97}.
Similarly, a growing body of evidence suggests that the LBV phenomenon includes 
2D or 3D processes. 
For example, \object{$\eta$~Car} is a binary \citep{dam00} and the Homunculus nebula 
around it is the most cited example of a bipolar outflow. 
The aspherical envelope of \object{$\eta$~Car} is not unique among LBVs;
\citet{gro06} found that \object{AG~Car} is a fast rotator, and hence may have a 
latitude dependent wind.  
The relevance of the binarity and fast rotation in the LBV phenomenon
in general is not yet clear and is the subject of vigorous research \citep[e.g.,][]{mar06, nie07}.

B[e] stars, or rather the stars that show B[e] phenomenon \citep{lam98}, 
are characterized by strong broad H Balmer emission lines (as are the classical Be stars) and
by the presence of narrow permitted and forbidden emission lines from low-ionization species,
like \ion{Fe}{II}, [\ion{Fe}{II}], or [\ion{O}{I}].
B[e] stars also show strong near/mid-IR excess that is attributed to hot cirumstellar dust.
Interested readers should refer to \citet{zic85, zic86, zic92, mir05} and 
references therein for further information on the B[e] phenomenon.
A particularly interesting subclass of the B[e] stars are the sgB[e]-s \citep{lam98}.
These are single B supergiants that show the B[e] phenomenon and are located near the 
Humphreys-Davidson limit \citep{hum79, hum84}. 
Their location on the HR diagram may suggest a link to the more massive LBVs or, 
alternatively, they may represent a second evolutionary path between Of and Wolf-Rayet (W-R) stars 
\citep{sch98, zic98}.

The envelope of sgB[e] stars is thought to be latitude dependent, with
a normal B supergiant wind at the pole and a dense and slowly moving equatorial 
flow \citep{zic85, zic86}.
Polarimetric observations by \citet{mag92, oud98, mel01} revealed that sgB[e]-s
have strong intrinsic polarization consistent with this picture.
The most widely accepted theory for the bi-modal nature of the envelope is the
rotationally induced bi-stability.
Around $T_{eff}$= 20,000 -- 25,000~K the characteristics of the line-driven wind 
abruptly change due to the recombination of \ion{Fe}{IV} to \ion{Fe}{III} \citep{pau90, lam91}.
If the sgB[e] stars are fast rotators, as suggested, the ensuing gravity-darkening can place the 
equatorial and polar regions on the opposite sides of the bistability, hence providing
a mechanism to produce the bi-modal envelope.
However, the bi-stability model is not without its weaknesses.
For example, gravity darkening reduces the flux at the equator, potentially inhibiting
any equatorial enhanced mass-flux. It is also difficult to maintain multiple scattering in the equator 
when the photons can easily escape in the polar direction.
(see discussion in \S\ref{section:disc}).
Consequently, other sgB[e] models have been proposed, like the presence of a Keplerian disk similar 
to those around classical Be stars. Further, disk winds as proposed by \citet{oud98}, the traditional 
wind compressed disk model of \citet{bjo93}, and even the magnetically confined disk models 
\citep{udd02,owo04} cannot be excluded. 
So far, theoretical efforts have primarily focused on understanding the hydrodynamic 
structure in the context of the bi-stability model, and trying to incorporate the bi-stability
jump into the \cite{CAK} formalism \citep[e.g.,][]{pel00, cur05}.

There have only been a few efforts to understand the ionization 
structure of the wind and to perform spectral analysis of B[e] stars.
Semi-analytical calculations of hydrogen and helium ionization 
were done by \cite{kra03} and \cite{kra06}, and they found the equatorial disk 
essentially neutral in these species all the way down to the stellar surface.
\cite{por03} attempted to reproduce the optical to infra-red continuum of
\object{R~126} (prototype sgB[e] in the LMC) by using bi-stability and 
Keplerian viscous disk models. Both of them reproduced
the optical and near-IR emission, but they failed to account for the dust emission
by an order of magnitude.
Recently, \cite{kra07} proposed that the dust and optical/near-IR continuum emission can be 
reconciled if the free-free and bound-free emission originates from the polar wind rather than
the equatorial disk.

A self-consistent spectral analysis of a stellar envelope needs to solve the coupled equations
for the radiation field and level populations. 
Obviously, this is a very difficult task in a non-LTE 2D or 3D envelope and 
compromises and simplifications are inevitable.
\citet{kra03} and \citet{kra06}, for example, used a simplified radiative transfer,
limited their analysis to optically thin or optically thick cases, and neglected ionization 
from excited levels; all of these can potentially have serious effects on the 
resulting ionization structure.
While significant insights can be gained by simple calculations, one needs 
numerical simulations for a fully self-consistent spectral analysis.

To provide a tool for such studies of Be/B[e] stars, binaries, and LBVs, we have 
been developing a code for axi-symmetric models.
So far, only basic tests and code verification were performed \citep{geo06, zsa06}, 
but we have now reached the point where scientifically meaningful simulations can be
performed.
The ionization structure of hydrogen and helium in axi-symmetric sgB[e] envelopes 
offers such an opportunity.  
The details of the program are discussed in \citet{geo06}
and \citet{zsa06} so we only briefly describe it in \S\ref{section:program}.
The hydrodynamic structures and the atomic models that are used in our simulations
are discussed in \S\ref{section:models}; computational issues are discussed in
\S\ref{section:issues}; and our results are presented in 
\S\ref{section:results}.
Simulated observations are shown in \S\ref{section:obs}, and
we draw our conclusions in \S\ref{section:disc} and \S\ref{section:sum}.

\section{The Code}\label{section:program}

Our C++ code, ASTAROTH, was developed for simulations of stellar envelopes.
However, it is flexible enough to be applied to any hot axi-symmetrical object with 
velocity gradients (including extragalactic objects, like AGN-s).  

The non-LTE level populations, the radiation field, and the electron temperature 
are calculated by simultaneously solving the equations of statistical equilibrium,
radiative transfer, and energy conservation. 
The short-characteristic method \citep[see e.g.,][]{mih78, kun88, bus00} is used 
to treat the continuum radiation transfer while bound-bound transitions are treated,
for simplicity, by the Sobolev approximation. 
The simultaneous solution of the equations is found by an approximate
lambda iteration \citep{ryb91, ryb92}.
The code has been tested by solving 2D pure scattering problems with grey opacity,  
as well as reproducing spherical symmetric models of CMFGEN \citep{hil98}, a well-established
code in stellar studies. 
In these tests, the new code reproduced the reference results within a few percent 
\citep[see,][]{zsa06, geo06}.

\section{Models}\label{section:models}

To simulate the hydrodynamical structure of a sgB[e] atmosphere we followed the 
approach of \citet{kra03}, but relaxed two of their simplifications. 
We used a $\beta$-law \citep{CAK} to describe the radial velocity and allowed for 
a varying electron temperature. 
Our velocity field was still simplified; only a latitude dependent radial velocity 
was included and the azimuthal and latitudinal velocities were set to zero.
To simulate the bi-modal wind, we described the radial velocity and the mass-loss 
per unit solid angle by
\begin{equation}\label{eq:1}
V_r(R,\theta)= V_{\infty}(\theta= 0 ) \cdot 10^{b_v \, sin^s \theta} 
\cdot \left( 1 - \frac{R_{*}}{R} \right) ^{\beta}
\end{equation}
and
\begin{equation}\label{eq:2}
\frac{\partial^2 M }{\partial t \partial \Omega}(\theta)= \frac{\partial^2 M}{\partial t \partial 
\Omega}(\theta= 0 ) \cdot 10^{b_m \, sin^s \theta} \; ,
\end{equation}
respectively.
The parameters $R$ and $\theta$ are the traditional polar coordinates, and $s$ controls 
the thickness of the equatorial disk; the larger it gets the thinner the disk.
In our simulations $s$ was set to 10, unlike in \citet{kra03} where 10 and 100 were used. 
In these initial calculations we avoided the higher values because of the need to have a much
finer spatial grid which would have substantially increased the computational effort.
In a future, detailed analysis of sgB[e]s, the effects of varying disk thickness will need to
be explored.

\begin{figure}
\centering
\includegraphics{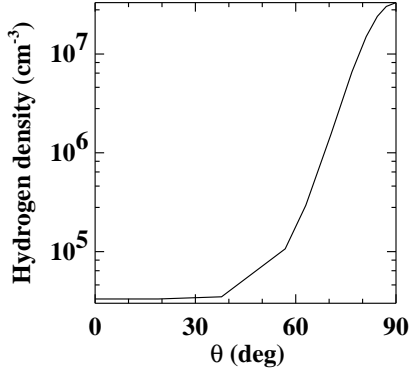}
\caption{
The density as a function of latitude at $R$=~100~$R_*$ in our models A and B (see Table~\ref{tab1}). 
Note that the density scale is logarithmic.
The density distribution for models C and D is similar in shape, but the values are 10 times
greater.
}  \label{fig:DB}
\end{figure}

\begin{figure}
\centering
\includegraphics[scale = 0.6]{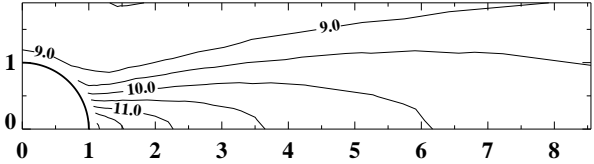}
\caption{
A meridional snapshot of the logarithmic hydrogen density near the star
for models A and B. 
The pole is toward the top of the page and the equator is horizontal. 
The envelope is axi-symmetric around the pole and top-bottom symmetric
around the equator. 
The units are in stellar radius and the thick circle at the center 
represents the surface of the star.
A similar snapshot for models C and D would look identical except with larger 
density values. 
Note, that the open contour lines near the stellar surface are artifacts of 
omitting the region R$<$1.1~R$_*$ from the plot which was done for clarity.
}  \label{fig2}
\end{figure}

Using Eq.~\ref{eq:2} and assuming top-bottom symmetry, the total mass-loss rate can be 
calculated by
\begin{equation}\label{eq:3}
\dot M = 4 \pi \cdot \int_{0}^{\frac{\pi}{2}} \frac{\partial^2 M}{\partial t \partial 
\Omega}(\theta) \cdot sin \theta \, d\theta \; .
\end{equation}
Also, because all non-radial velocities are zero, the gas density in the wind,
$\rho (r, \theta )$ is given by
\begin{equation}\label{eq:4}
\rho (R,\theta) = \frac{\partial^2 M}{\partial t \partial \Omega}(\theta) \cdot \frac{1}{R^2 
V_r(R,\theta)} \; .
\end{equation} 
In our models we used b$_v$ = $-2$ and b$_m$= $1$ which resulted in a 3 order of magnitude 
density enhancement and a 2 order of magnitude velocity decrease between the pole and the disk 
(see \S\ref{section:disc} for discussion on the validity of Sobolev approximation in the disk).
The 2D density structure is displayed in Fig~\ref{fig:DB} and Fig~\ref{fig2}.
The bi-modal structure of the wind is apparent with 3 order of magnitude difference between 
the pole ($\theta$= 0$^{\circ}$) and the equator ($\theta$=~90$^{\circ}$).

There is a significant difference between our approach and that of \cite{kra03}.
They did not deal with the photosphere, and used \citet{kur79} spectra
to simulate the stellar radiation field.
ASTAROTH, similarly to CMFGEN, solves for the unknown level populations and ionization
structure, and calculates spectra for both the photosphere and the wind.
It uses the same methods as CMFGEN \citep{hil98} to create an approximate hydrostatic photosphere.
In brief, the wind follows a simple beta-like velocity law \citep{CAK}, and for this work starts at a radius 
$R_{phot}$ defined by $V_r(R_{phot},\theta)$= 1~km~s$^{-1}$.
Below this radius the density follows an exponential law and the velocity is set by the continuity 
equation for the given mass-loss rate.
The value of $R_{phot}$ is latitude dependent in the models presented here, but it was 
always nearly $R_*$ ($<$1-2\% difference).

For the stellar parameters, listed in Table~\ref{tab1}, we chose values that are broadly representative of 
the most luminous sgB[e]-s \citep[see,][]{zic85, zic86} and some less luminous LBVs 
\citep[e.g., \object{P~Cygni},][]{pau90, dre85}.
The mass-loss range covers the upper end of the range used in \citet{kra03} and includes their 
baseline models \cite[models A, C, and F in][]{kra03}.
The suspected mass-loss rate of \object{R~126} \citep[$\sim$4$\times$10$^{-5}$~M$_{\odot}$~yr$^{-1}$,][]{bjo98} 
is also within our mass-loss range.
We vary the effective temperature between T$_{eff}\sim$22,500~K (L= 1.5$\times$10$^6$~L$_{\odot}$), that of \object {R~126}, 
to T$_{eff} \sim$ 18,000~K (L=6$\times$10$^5$~L$_{\odot}$) which approximately covers the sgB[e] temperature 
range in the Magellanic Clouds \citep[see,][]{zic85, zic86}. 
 
Finally, the atomic model is presented in Table~\ref{tab2}.
We used essentially H/He atmospheres with a fairly large number of levels included.
A few levels and ionization states of carbon were also included to achieve more realistic 
heating and cooling terms. The fractional (number) abundances of He and C was 0.1 and 
9.82$\times$10$^{-4}$, respectively. The later number represents the net
abundance of CNO elements.

\begin{table}
\caption{Description of the Models\label{tab1}}

\begin{tabular}{lcccc} \hline\hline
Model                        &   A   &   B    &   C   &   D    \\
M$_*$                        & \multicolumn{4}{c}{30 M$_{\sun}$}  \\
R$_*$                        & \multicolumn{4}{c}{82 R$_{\sun}$}   \\
T$_{eff}$                    & 22500~K & 18000~K & 22500~K & 18000~K \\
$\frac{\partial^2 M}{\partial t \partial \Omega}$($\theta$=0) & \multicolumn{2}{c}{8$\times$10$^{-8}$~$\frac{M_{\odot}}{yr ~ str}$} & \multicolumn{2}{c}{8$\times$10$^{-7}$~$\frac{M_{\odot}}{yr ~ str}$} \\
 & & & & \\
$\frac{\partial^2 M}{\partial t \partial \Omega}$($\theta$=$\frac{\pi}{2}$) & \multicolumn{2}{c}{8$\times$10$^{-7}$~$\frac{M_{\odot}}{yr ~ str}$} & \multicolumn{2}{c}{8$\times$10$^{-6}$~$\frac{M_{\odot}}{yr ~ str}$} \\
 & & & & \\
$\dot M$         & \multicolumn{2}{c}{3.3$\times$10$^{-6}$~M$_{\odot}$~yr$^{-1}$} & \multicolumn{2}{c}{3.3$\times$10$^{-5}$~M$_{\odot}$~yr$^{-1}$} \\
V$_{\infty}$($\theta$=0) & \multicolumn{4}{c}{2000~km~s$^{-1}$} \\
V$_{\infty}$($\theta$= $\frac{\pi}{2}$) & \multicolumn{4}{c}{20~km~s$^{-1}$} \\
$\beta^a$                      & \multicolumn{4}{c}{0.8}   \\ \hline
\end{tabular}

$^a$ -- Power for the standard $\beta$ velocity law \citep{CAK}.
\end{table}

\begin{table}
\caption{Atomic Model\label{tab2}}
\begin{tabular}{lllc} \hline\hline
Specie       &~&~&  Number of Levels \\
\ion{H}{I}   &~&~&   20  \\
\ion{H}{II}   &~&~&    1  \\
\ion{He}{I}  &~&~&   11  \\
\ion{He}{II}  &~&~&   20  \\
\ion{He}{III}  &~&~&   1  \\
\ion{C}{II}   &~&~&    9  \\
\ion{C}{III}   &~&~&   10  \\
\ion{C}{IV}   &~&~&    5  \\
\ion{C}{V}   &~&~&    1  \\ \hline
\end{tabular}
\end{table}

\section{Computational Issues}\label{section:issues}

Previous works on stellar winds and sgB[e] stars \citep[e.g.,][]{kra03} have found that ionization changes can 
occur suddenly, in the form of ionization fronts.
One of our main concerns was, therefore, whether our spatial grid would adequately resolve
a potential 2D ionization front in our simulations.
The standard grid included 60-80 depth points and $\sim$10 latitudes, both of them spaced semi-irregularly with the 
limitation that all latitude must have the same radial grid.
The only problem we encountered was the sharp hydrogen recombination front that occurred in model D, and the spatial 
grid had to be adjusted by hand to better resolve the front (adaptive gridding capability is not yet included in 
ASTAROTH).
Otherwise the standard grid was adequate for our models.
Simulations on a denser grid with 120 depth and 20 latitude points revealed no qualitative changes in the 
populations and temperatures.

The convergence was also sometimes slowed by oscillations in the low radiation field/high density 
equatorial disk.
As the statistical equilibrium and energy conservation equations are highly nonlinear in the populations, 
radiation field, and electron temperature, solution techniques can provide oscillating solutions. 
Often the oscillations arose at depths, and in populations, where changes would have a negligible
influence on the emergent spectrum. Unfortunately, since we currently use a stopping criterion based on the maximum population change, such oscillations can greatly increase the total computational effort. 
To dampen the oscillations we linearized the statistical equilibrium equations with populations averaged
over several previous iteration cycles. 
Switching to linear interpolations of quantities, instead of the standard cubic or 
parabolic approximations, also improved the convergence. 
The linear interpolation is an extremely well-behaved and stable approximation and it would be the preferred 
method if it provided the necessary accuracy.
Unfortunately, this is not the case for spatial grids that are practical for ASTAROTH simulations; therefore,
linear interpolation was limited to grid-points immediately around the trouble spot.

\section{Ionization and Temperature Structure}\label{section:results}
 
Figs.~\ref{fig3}, \ref{fig4}, and \ref{fig5} show the electron temperature, hydrogen,
and helium ionization structures, respectively, in our model envelopes. 

\subsection{Electron Temperature}

\begin{figure}
\centering
\includegraphics[scale = 0.6]{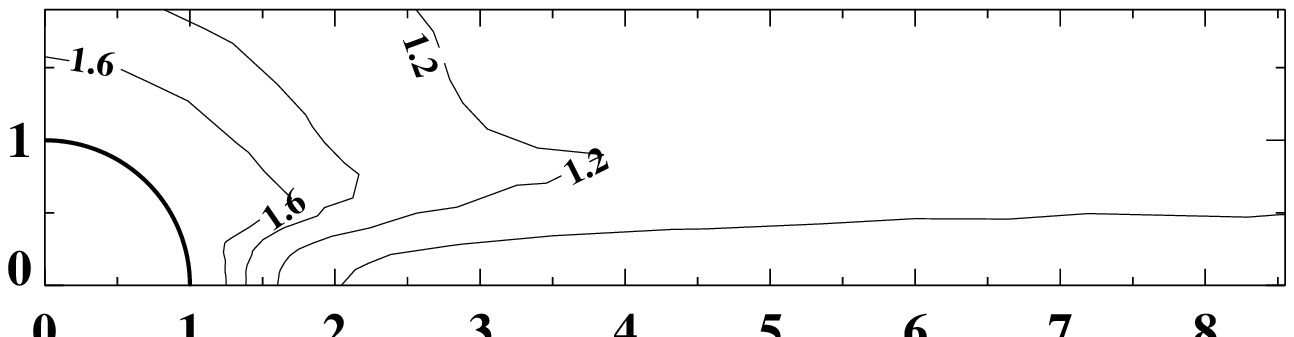} 

~~\newline
\includegraphics[scale = 0.6]{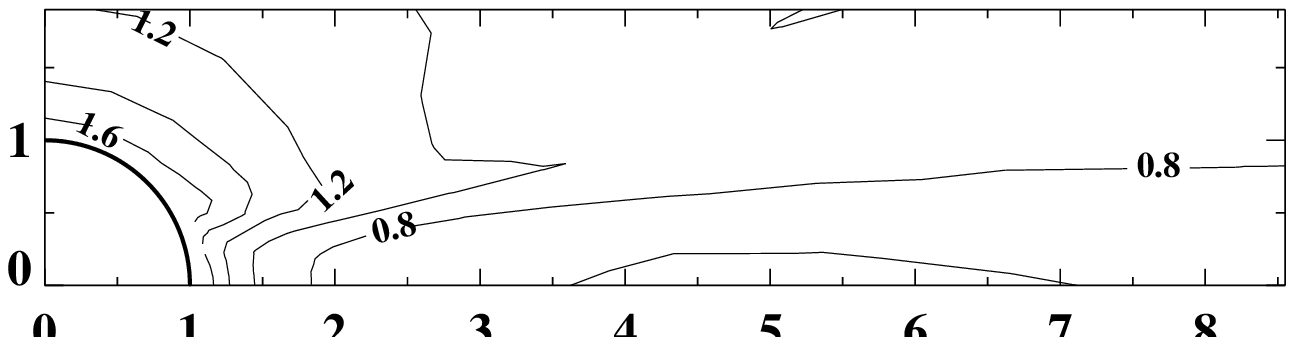}

~~\newline
\includegraphics[scale = 0.6]{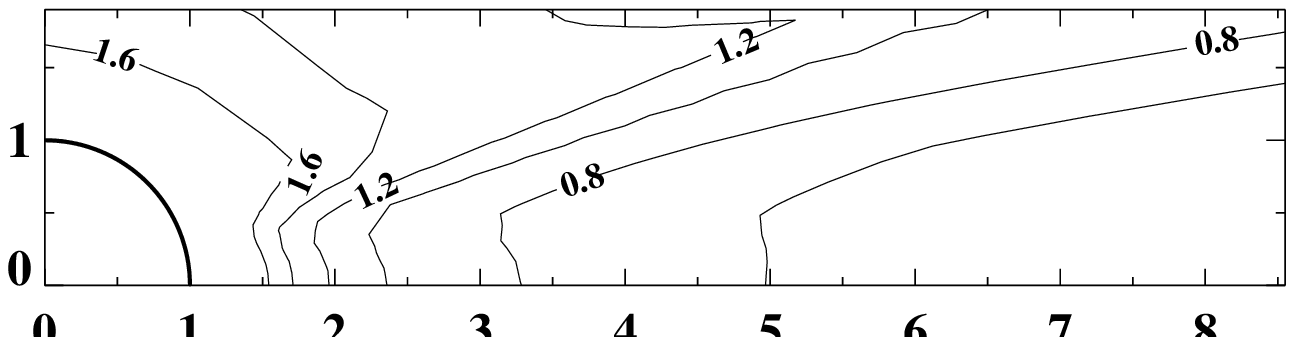}

~~\newline
\includegraphics[scale = 0.6]{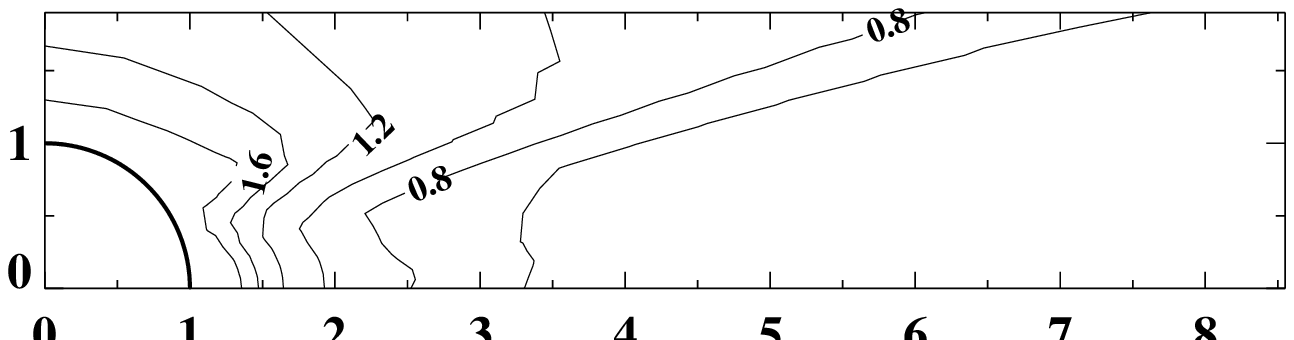}
\caption{
The temperature structure near the star in models A, B, C, and D (top to bottom).
The layout of the plots is the same as that of Fig.~\ref{fig2}. 
The contour levels are from 6,000~K to 16,000~K in 2,000~K increments and shown in 
10,000~K units.
}  \label{fig3}
\end{figure}

The electron temperature varies between $\sim$100,000~K to $\sim$20,000~K in the inner hydrostatic 
atmosphere (not shown in Fig.~\ref{fig3}), and it was a few 1,000~K in the outer envelope.
The behavior of $T_e$ in the wind is similar, as expected, for all models --- the equator is 
cooler than the pole at the same radius. 
The effect of the lower luminosity in models B and D is a shift 
of the overall structure closer to the stellar surface (see Fig~\ref{fig3}), while
increasing the mass-loss results in a thicker cool disk.
It is immediately obvious from Fig~\ref{fig3} that models C and D have a bulkier and wider
``disk" than models A and B.

The temperature in our models is too hot for dust formation.
Nowhere in our models the temperature falls below $\sim$1,500~K which is the upper limit
for dust formation \citep{por03}. 
We cannot conclude, however, that dust is formed beyond $R > 100-200 R_*$ (the outer boundary
of our models) because of the relative simplicity of our models and the neglect of 
adiabatic cooling in our simulations.
We will address this question in follow-up studies with models constrained
by the observations of individual stars.

\subsection{Hydrogen Ionization}

\begin{figure}
\centering
\includegraphics[scale = 0.6]{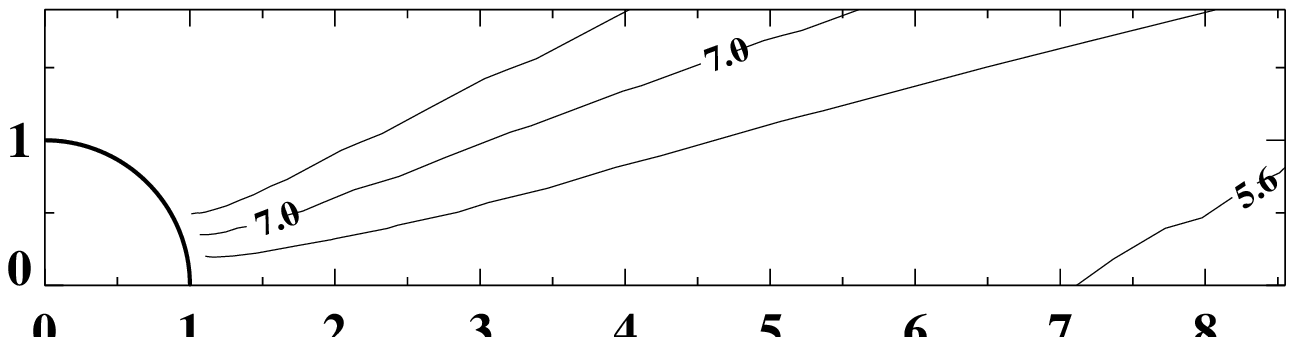}

~~\newline
\includegraphics[scale = 0.6]{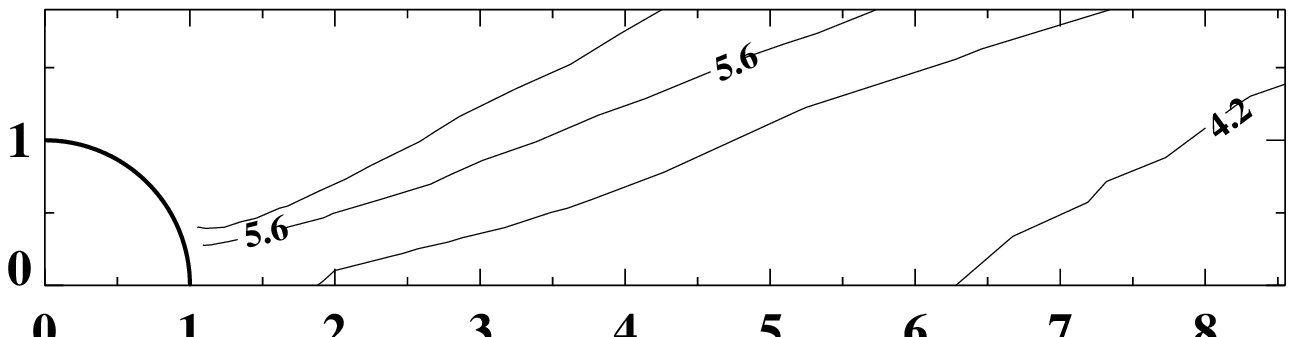}

~~\newline
\includegraphics[scale = 0.6]{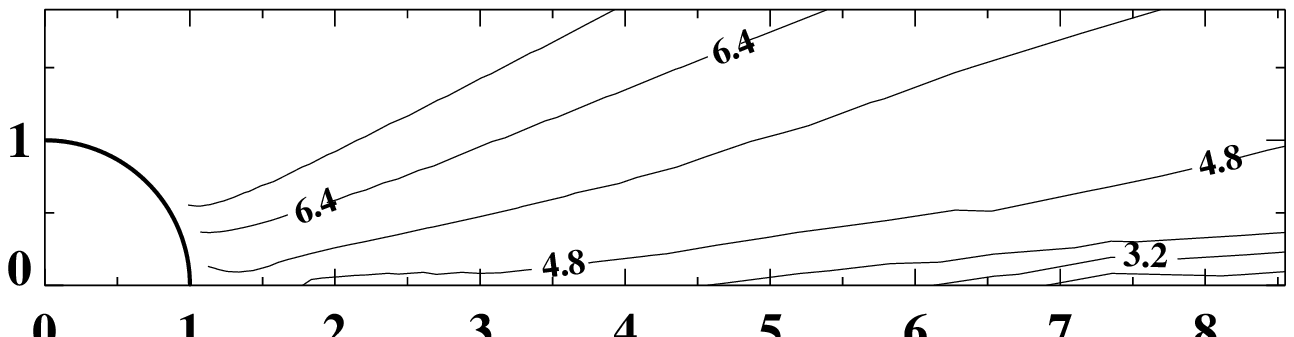}

~~\newline
\includegraphics[scale = 0.6]{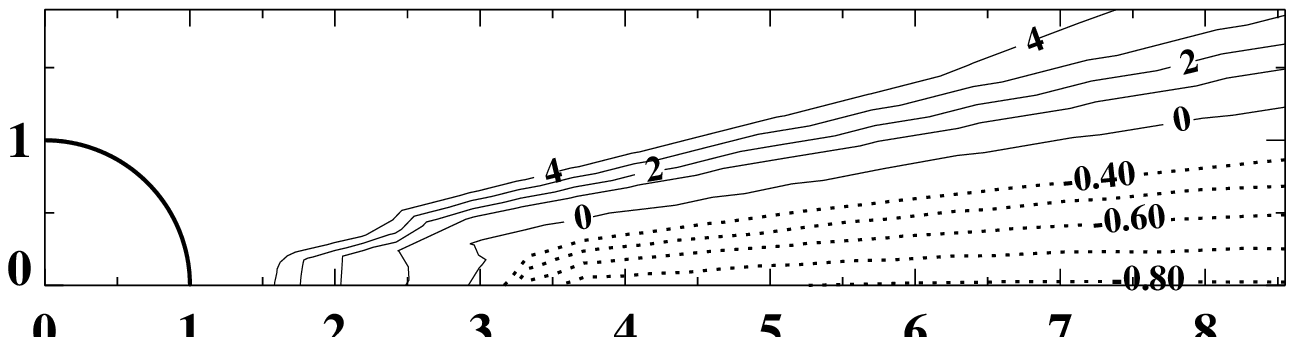}
\caption{
The $log \left( \frac{N_{H~II}}{N_{H~I}} \right)$ contours for models A, B, C, and D
(top to bottom). 
The layout of the figures is the same as that of Fig.~\ref{fig2}.
The thin solid lines represent ratios above 0 (ionized H) and the thick dotted lines 
show neutral regions.
Note, that the lower the ratio the more neutral the hydrogen is!
The spacing between levels is different from figure to figure and between neutral and 
ionized regions.
}  \label{fig4}
\end{figure}

Fig.~\ref{fig4} shows the ionized to neutral H ratio in logarithmic scale where neutral H
means the total population of all \ion{H}{I} levels. 
It was very difficult to display the wide range of ionization levels occurring in our models.
Since there were orders of magnitude differences even within a single model, we opted for using a 
separate set of contour levels for each model.

The hydrogen ionization shows latitudinal variations similar to those of the temperature.
The equatorial region is more neutral than the pole in all models, and the
lower the effective temperature  or higher the mass-loss the more neutral the inner envelope.
In only one model, model D, does a neutral H disk form, and then only beyond
$\sim$3$R_*$.
This contradicts the result of \citet{kra03} who found predominantly 
neutral hydrogen disks, even nearly at the surface, for very similar models.
In our models A, B, and C the neutral hydrogen is negligible compared to \ion{H}{II} (inside 10$R_*$),
although there is a tendency for the neutral hydrogen fraction to rise at larger radii.
The high level of ionization prevails even if the effective temperature is lowered (model B) or the mass-loss 
rate is increased (model C).
Only the combined effect of the two (model D) produced a neutral hydrogen disk.
We will further discuss the differences between our results and those of \citet{kra03}
in \S\ref{section:disc}.

\subsection{Helium Ionization}

\begin{figure}
\centering
\includegraphics[scale = 0.6]{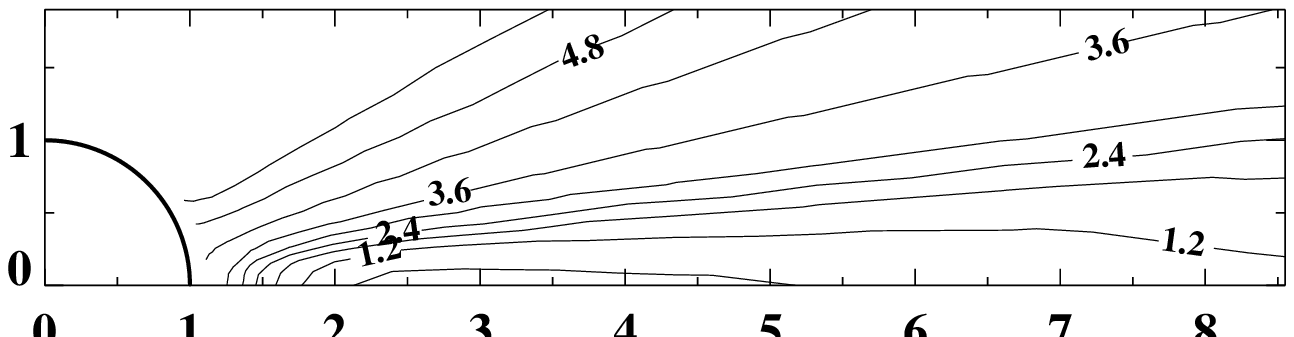}

~~\newline
\includegraphics[scale = 0.6]{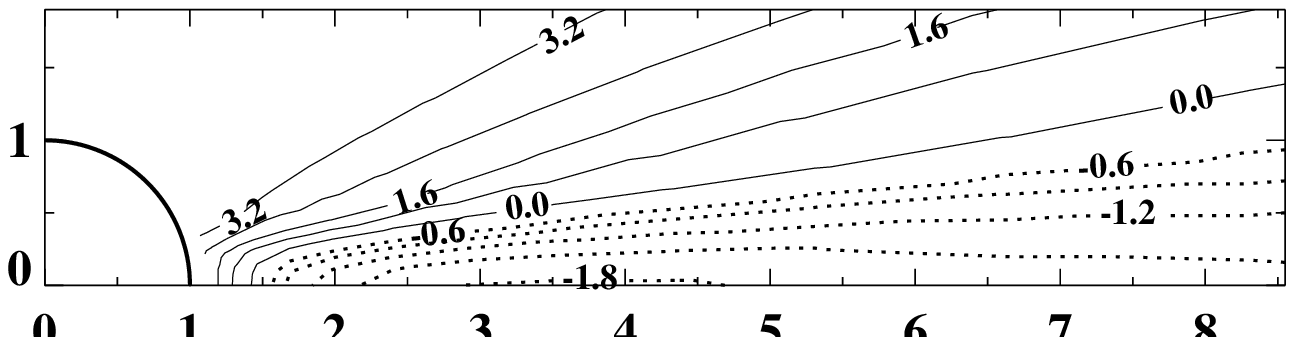}

~~\newline
\includegraphics[scale = 0.6]{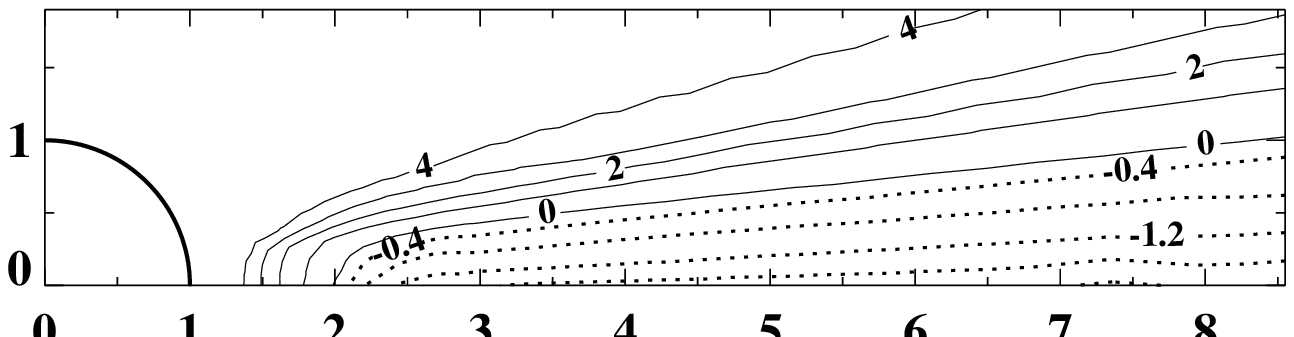}

~~\newline
\includegraphics[scale = 0.6]{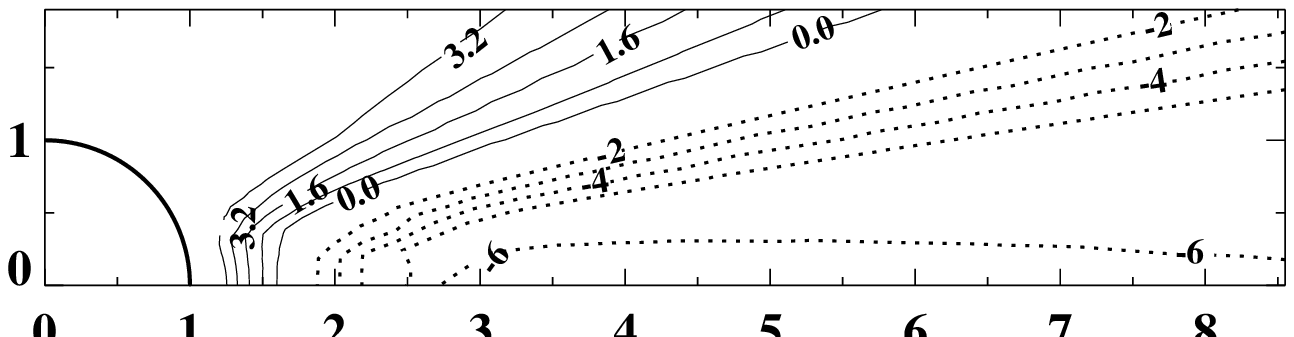}
\caption{
The  $log \left( \frac{N_{He~III} + N_{He~II}}{N_{He~I}} \right)$ contours for models A, B, C, and D 
(top to bottom). 
The layout of the plots and the definitions of the contour lines are the same as 
those of Fig.~\ref{fig2} and \ref{fig4}, respectively.
}  \label{fig5}
\end{figure}

Fig.~\ref{fig5} shows the logarithm of the ionized to neutral He ratio.
Ionized He means the total population of all \ion{He}{II} and \ion{He}{III} levels. 
The ionization level of helium is markedly lower than that of hydrogen, a result to
be expected given the low effective temperatures which were adopted.
As opposed to H, He forms a neutral disk in all but one of our models --- 
in model A, only an enhancement of neutral helium occurs in the equatorial region.
The neutral disks show characteristics similar to those of the 
temperature distribution or hydrogen ionization; e.g,
lowering T$_{eff}$ results in an inward shift of the neutral regions as seen in the
last two panels of Fig.~\ref{fig5}. 
Higher mass-loss causes the neutral disk to become thicker.
It also appears that the thicker the disk, the sharper its ionization boundaries.
The neutral He disk reaches down deeper than the hydrogen disk in model D, but
it is still truncated at $\sim$2~$R_*$. 
  
\section{Observed Profiles}\label{section:obs}

Observed profiles for a converged ASTAROTH model are calculated independently by an auxiliary 
routine \citep[described in][]{bus05}. 
We also used a Monte-Carlo code \citep{hil91} for polarization calculations, and to verify 
the results of the auxiliary routine.
Below we discuss the properties of the observed H$\alpha$ profile, which arises from
the disk,  and the  \ion{C}{IV}~$\lambda\lambda$1550 doublet profile, which primarily arises in the polar wind. 
We also discuss similarities and differences between the computed lines, and those seen for
our benchmark B[e] supergiant \object{R~126}, although we stress that this work is not a spectral analysis 
of this star.
For the calculations presented in the following sections we generally adopted 10\,km\,s$^{-1}$
for the Doppler parameter of the intrinsic line absorption/emission profile. Because of the low
velocities in the disk, the adopted Doppler parameter can have a substantial influence on
the line profile shape, and model profiles calculated with Sobolev approximation are
significantly different.
Furthermore, as the azimuthal velocities are zero in our models, the 
profiles are not rotationally broadened.

\subsection{H Line Profiles}

\begin{figure}
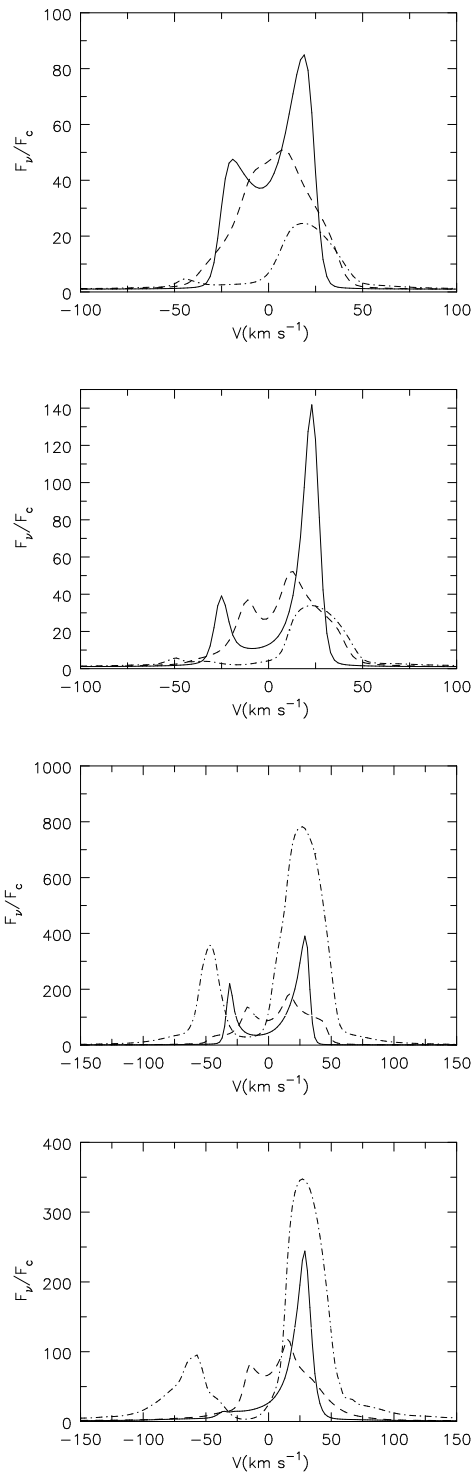

\includegraphics[angle=270, scale=0.25]{8293fg15.ps} \newline  
\includegraphics[angle=270, scale=0.25]{8293fg16.ps} \newline
\includegraphics[angle=270, scale=0.25]{8293fg17.ps} \newline
\includegraphics[angle=270, scale=0.25]{8293fg18.ps}
\caption{
Simulated observations of H$\alpha$ for models A, B, C, and D (top to bottom).
There are three viewing angles, $i$= 0$^{\circ}$ (pole-on, solid line), 45$^{\circ}$ (dashed line), and 
90$^{\circ}$ (edge-on, dash-dotted line).
Note that all profiles are rectified.
}  \label{fighyd}
\end{figure}

The shape of the H$\alpha$ lines in Fig.~\ref{fighyd} are very sensitive to the observer's viewing
inclination~($i$). Depending on the model either a single or double emission peak is seen; sometimes
the emission peaks are well resolved, while at other times they blend into a continuous emission feature.
For $i=0$, the half-widths at the line base (HWLB) are 30--50~km~s$^{-1}$ which is slightly larger than 
$V_{\infty}$ in  the equatorial plane (20~km~s$^{-1}$). Very weak wings, arising from emission in
the wind outside of the equatorial latitudes, is also present. At higher inclinations the profiles are broader,
and generally the emission is red-shifted. The red-shifted peak is always stronger than the blue-shifted peak,
a characteristic feature often seen in real sgB[e] observations \citep[see e.g.,][]{zic85, zic86}.

The double-peak profiles arise from optical depth effects in our simulations, and are not the result
of rotation. 
The keys to understanding the double-peak profile are (1) that  
H$\alpha$ is optically thick at the line center in all four models, and hence photons generally escape 
in the line wings where $\tau \sim$~1, and (2) that the Doppler velocity is comparable to the disk 
outflow velocity. 
Let's consider a spherical model with an extended atmosphere but no velocity field. 
In general, the source function will decrease with radius, and thus
the maximum of the line source function, along a line of sight parallel to z-axis, will be in 
the xy plane.
Optical depth unity will occur in the xy plane at two wavelengths, symmetrically located about line center. 
This will produce a distinct double-peaked profile with, in the absence of a velocity field,
both peaks being of equal strength. However, when a velocity field is present, the red peak
will be stronger than the blue. Photons emitted on the red side only see material
moving away, and never come into resonance. On the other hand, photons emitted in the blue side
must traverse material that is resonant with it, thus reducing the line intensity. 
The separation of the two peaks depends on the line width, and the overall profile and 
equivalent width is very sensitive to the adopted Doppler and
wind velocities. When the wind velocity is much larger than the Doppler velocity we see a classic
wind emission line with an optical-dependent redshift of 1 to 3 Doppler widths.
As apparent from the above discussion, the blue/red peak asymmetry is not a specific feature of 2D models 
and also works for spherical envelopes, provided that
the line is optically thick and the Doppler and flow velocities are comparable.

The shape of the pole-on viewed H$\alpha$ in our model A is very similar to that 
seen in the spectra of \object{R~126} \citep[see Fig.3 of][]{zic85}. 
However, the simulated profile lacks the broad wings of the observations and is a 
factor of $\sim$10 weaker than it should be.
This indicates that the density is much lower in Model A than it is in the envelope of \object{R~126},
and highlights the importance of frequency redistribution in electron scattering (the profiles
in Fig.~\ref{fighyd} were calculated ignoring electron-scattered line photons).
The higher mass-loss rate in models C and D increases the equivalent width (it still falls short by a factor of $\sim$4,
except at $i$= 90$^{\circ}$), but the profiles now are very different from that of \object{R~126}.
The very strong absorption around and blueward of the line center is not observed.
This central absorption is very prominent for models B, C, D, and may indicate that 
hydrogen is more neutral in the line forming region of these models than it is around \object{R~126}.

Table~\ref{tab3} lists the equivalent widths (EW), the continuum levels across H$\alpha$, the total intensities
arising in the H$\alpha$ profiles (line fluxes), and the H$\alpha$ to H$\beta$ line  flux ratios.
The equivalent width varies with viewing angle, although the variation is weak for low to moderate inclinations.
The continuum flux also varies with inclination angle. In models D and C the variation in continuum flux
is large (the continuum is $\sim$10$\times$ smaller for the $i=90^{\circ}$ model than it is for $i=45^{\circ}$), and 
this causes the EW to be higher for the $i=90^{\circ}$ models, despite the lower line flux.

In modeling of O stars, LBVs, and  W-R stars simple scaling laws can provide valuable insights into
wind spectra, and greatly facilitate modeling through a reduction in the parameter space that
needs to be explored. For W-R stars, \cite{sch89} found that models with the same effective
temperature, and the same ``optical depth invariant''  parameter, 
$Q = \frac{\dot M}{\left( v_{\infty} R_* \right) ^{1.5} }$, showed similar spectra. The
same invariant also holds for recombination dominant transitions in O stars \citep{pul96, kot98}.
With the added complexity of 2D models, the identification of useful invariants is not so easy.

The extra difficulty of the 2D models is apparent from Table~\ref{tab3}. 
First, continuum fluxes, line fluxes, and line EWs vary with the observer's inclination. 
Thus knowing the observer's inclination is crucial if meaningful 
envelope parameters are to be derived from the spectra.
It is also apparent that for B[e] stars ionization effects are important, even for hydrogen. 
Models C and D, for example, have identical $Qs$ but different EWs and line fluxes.
Another complexity, not seen in W-R models, is that the EW and line-flux of H$\alpha$ depends on the 
adopted Doppler velocity.
When spectra for the very same models but with $V_{turb}$=~5~km~s$^{-1}$ were 
calculated (not shown), the EWs and line fluxes decreased by $\sim$40\%.
The H$\alpha$ to H$\beta$ ratio, on the other hand, was not  affected by the lower turbulent velocity.

To fully explore the effects of parameters on the observables
is beyond the scope of this paper but simple trends can already be identified. 
It is striking, for example, that the ratio of the continuum intensities for $i$=~45$^{\circ}$ and 0$^{\circ}$, 
especially in models C and D, is close to cos(45$^{\circ}$)=~0.71 which is also
the ratio of the projected area of a disk to its total area when tilted by 45$^{\circ}$.
One would expect such ratios if a substantial portion of the continuum around H$\alpha$ 
formed in the transition region between the polar wind and the dense disk. 
The visible area of this disk ``surface'' would then control the continuum level.
Furthermore, most of the H$\alpha$ emission has to originate from the equatorial region 
since many of its characteristics cannot be explained in the context of a nearly spherical
polar wind; e. g., its sensitivity to the observer's inclination angle.   
Simulations with CMFGEN (performed by the authors) also showed that one would need 
extreme stellar parameters to reproduce observed sgB[e] features in spherical models.
Mass-loss rates in excess of 10$^{-4}$~M$_{\odot}$~yr$^{-1}$ and
 $v_{\infty} \le$ 200~km~s$^{-1}$ are required 
to match the observed linewidth and EW \citep[$\sim$900~\AA\ ,][]{zic85} for R~126.   

% This version of the table was calculated by Vturb= 10 km/s.
\begin{table}
\caption{H$\alpha$ Equivalent Widths and Line Fluxes \label{tab3}}
\begin{tabular}{lccccc} \hline\hline
Model  &   $i$~$^I$   & $F_c$~$^{II}$ & EW (\AA ) & $F_{l}$ = EW$\times F_c$ &   
$\frac{F_{l} (H \alpha )}{ F_{l} (H \beta )}$  \\
 A     &              &               &           &              &           \\
       &  0$^{\circ}$ &     1060      &     61    &     64681    &     3.8   \\
       & 45$^{\circ}$ &      870      &     53    &     45736    &     3.8   \\
       & 90$^{\circ}$ &      384      &     21    &      8198    &     5.3   \\
 B     &              &               &           &              &           \\
       &  0$^{\circ}$ &      754      &     54    &     40754    &     4.3   \\
       & 45$^{\circ}$ &      615      &     49    &     30320    &     4.4   \\
       & 90$^{\circ}$ &      267      &     30    &      8013    &     7.1   \\
 C     &              &               &           &              &           \\
       &  0$^{\circ}$ &     3177      &    195    &    618689    &     5.7   \\
       & 45$^{\circ}$ &     2263      &    204    &    461810    &     5.7   \\
       & 90$^{\circ}$ &      107      &    873    &     93375    &     6.4   \\
 D     &              &               &           &              &           \\
       &  0$^{\circ}$ &     1725      &    116    &    200859    &     6.4   \\
       & 45$^{\circ}$ &     1195      &    134    &    160345    &     6.6   \\
       & 90$^{\circ}$ &      134      &    392    &     52511    &     7.1   \\ \hline
\end{tabular}

Note that the values (except the ratios) in this table are \\
rounded up to the closest integers!

$^I$ : Viewing angle, $i$= 0$^{\circ}$ (pole-on), 45$^{\circ}$, 90$^{\circ}$ (edge-on).

$^{II}$: Continuum intensity across H$\alpha$ (in Jansky). Constant to 1-2~\%.
\end{table}

It is more complicated to understand the spectra when the disk is viewed edge on.
At $i$= 90$^{\circ}$ the continuum is very low because the stellar radiation is seen through the dense 
equatorial disk and the projected area of the disk is small. The very large EWs for $i$= 90$^{\circ}$ 
viewing angle 
(see Table~\ref{tab3}) are principally the result  of the disproportionately low continuum; 
for high inclination angles the continuum weakens more with inclination angle
than does the H$\alpha$ emission. 
Since the continuum is low at every wavelength, these stars may appear unusually faint,
and thus their luminosities and mass are underestimated.

A further important feature of the H$\alpha$ emission is its T$_{eff}$ dependence 
which mainly affects the absorption at and blueward of the line center.
The lower the effective temperature is the stronger the absorption (see, for example, the difference between models 
B and A in Fig.~\ref{fighyd}).
This is also apparent on the lower equivalent widths for models B or D relative to their respective
high T$_{eff}$ counterparts.
It is not surprising that the hydrogen in these models is also less ionized and excited (relative to the
models with higher T$_{eff}$) which is clearly shown by the H$\alpha$ to H$\beta$ flux ratios listed in 
Table~\ref{tab3}. 
The larger population in the low-lying states can also increase the self-absorption
in the Balmer lines; hence it can explain the stronger absorption in models B and D. 

\begin{figure}
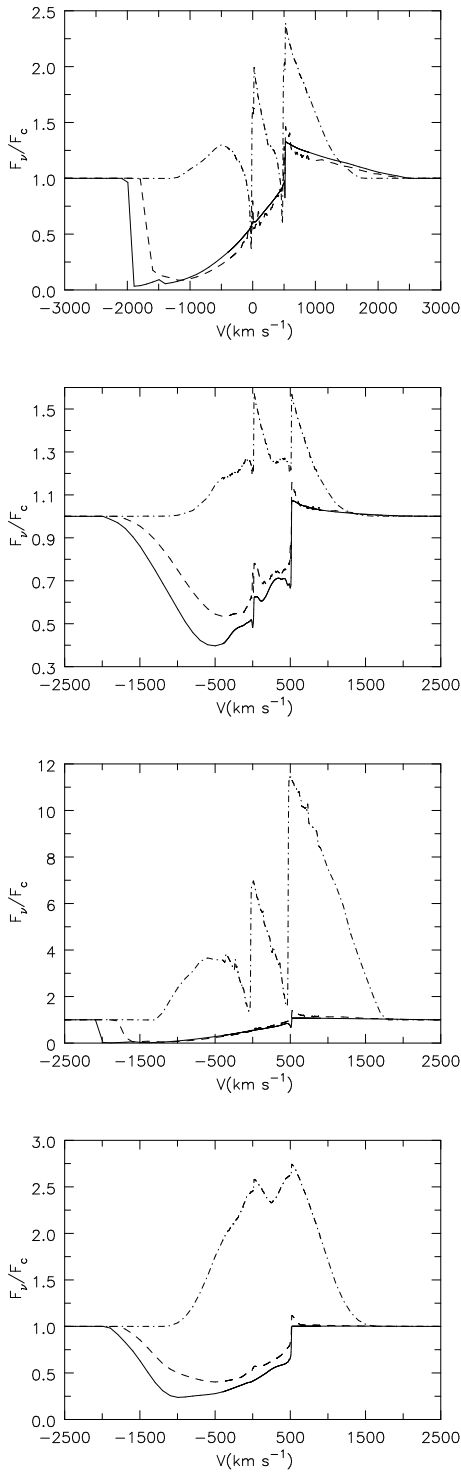

\includegraphics[angle=270, scale=0.25]{8293fg19.ps} \newline
\includegraphics[angle=270, scale=0.25]{8293fg20.ps} \newline
\includegraphics[angle=270, scale=0.25]{8293fg21.ps} \newline
\includegraphics[angle=270, scale=0.25]{8293fg22.ps}
\caption{
Same as Fig.~\ref{fighyd} but for \ion{C}{IV}~$\lambda\lambda$1550. 
Note that the velocity scale is centered on the doublet line with shorter wavelength.
}  \label{figciv}
\end{figure}

\subsection{\ion{C}{iv} Profile}

The \ion{C}{iv} doublet profile, for models A, B, C, \& D, and for $i$=0$^{\circ}$, 45$^{\circ}$, \& 90$^\circ$
is shown in Fig.~\ref{figciv}. A striking feature of the observed profiles in all models is the lack of
red-shifted emission at pole-on and intermediate viewing angles (i.e., $i$=0$^{\circ}$ and 45$^{\circ}$).
This is also one of the notable features in the International Ultraviolet Explorer (IUE) observations of \object{R~126} \citep[Fig.~5 in][]{zic85}. 

The behavior of \ion{C}{IV} is the exact opposite of what we saw in H$\alpha$.
\ion{C}{IV} is mainly in absorption formed in the polar wind, and shows little sensitivity to the viewing angle 
(at least until very nearly edge-on view).
This is because the polar wind occupies a large solid angle around the star and it is also nearly spherical.
A viewing angle of 0$^{\circ}$ or 45$^{\circ}$ makes little difference, apart from the position of the blue 
absorption edge.
The role of the dense equatorial region in the formation of the \ion{C}{IV} lines is to block the emission from 
the far side of the envelope.
The profiles in Fig.~\ref{figciv} are also very sensitive to the effective temperature of the models.
The absorption is much weaker for the lower T$_{eff}$ models (B and D) than for their high temperature
counterparts (A and C).

The edge-on viewed spectra in Fig.~\ref{figciv} are again complex.
There is always a broad and strong emission component, the signature of the polar 
wind.
They appear to be very strong in the rectified spectra only because they 
are seen against a very weak continuum. 
In three of the predicted profiles, doublet absorption components can also be seen. 
Their strength and shape is highly model dependent, and
presumably arises from ``disk'' absorption of both line and continuum photons.
The \ion{C}{IV} profiles potentially offer an opportunity to constrain the
thickness of the equatorial disk if the star is viewed nearly edge-on.

\section{Discussion}\label{section:disc}

Obviously the substantial discrepancy between our ionization results and those of
\citet{kra03} needs an explanation. 
We think that the nebular approximation used by 
\cite{kra03}, is the reason for the differing results. 
In the nebular approximation it is assumed that all recombination to n$>$1 levels 
of hydrogen will cascade down to level n=1 and the photo-ionizations from the n$>$1 levels are completely negligible. 
In dense envelopes this is unlikely to be the case, as has been seen in studies of LBVs, W-R, and O stars. 
\cite{dre85} studied \object{P~Cygni}, an LBV with a mass-loss rate and luminosity similar
to those of our models, and found that photo-ionization from the n$>$1 states were crucial in determining 
the hydrogen ionization level.
Similarly, \cite{pau87} recognized that ionization from excited states shifts the ionization level of elements higher 
in the wind of \object{$\zeta$~Pup} (O4f), and \cite{hil83} and \cite{hil87} found that ionization from the $n=2$ state of He was crucial in maintaining the He ionization balance in WR winds.

\begin{figure}
\centering
\includegraphics[angle=0, scale= 1.0]{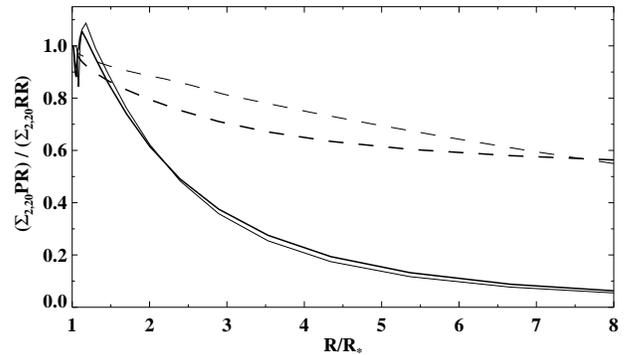}
\caption{
The ratio of the total H photo-ionization (PR) from levels with n$>$1 to the
total recombination (RR) to levels with n$>$1 for the pole (solid line) and
equator (dashed line) of model A. The thick lines are for ASTAROTH (the
actual 2D model) and the thin curves are the CMFGEN results (spherical
models created for the given latitude). 
}  \label{fig:PRRRrat}
\end{figure}

During a CMFGEN or ASTAROTH simulation, the radiative recombination to and photo-ionization 
from all included levels can be monitored, so their importance can be easily assessed.
Figure~\ref{fig:PRRRrat} shows the ratio of the total photo-ionization from the excited states of 
hydrogen to the total recombination to the same levels for both CMFGEN and ASTAROTH models.
For the nebular approximation to be valid, the ratios in Fig.~\ref{fig:PRRRrat} need to be very small.
However, the figure shows the ratios are significant, especially for the equatorial regions; therefore,
we can conclude that the nebular approximation is not valid in the inner envelopes of these models.
Ionization from the first few excited levels is extremely important, especially in the disk.

To simulate the nebular approximation with ASTAROTH is not trivial.
One needs to shut down the photo-ionization terms from excited states in 
the less dense regions, but needs to keep them in the hydrostatic photosphere to
recover LTE.
Nevertheless, we experimented with spherical ASTAROTH models of the equatorial region of
model A.
To do this, we had to alter a few features of ASTAROTH.
First, the inner boundary condition was changed from the diffusion approximation to
a black-body function with T$_{eff}$= 22,500~K.
With this modification the inner hydrostatic envelope could be omitted and the radiative
transition rates could be ``safely" modified to simulate the nebular approximation.  
When we ran the above simulations (equatorial region of model A) with this version of the code,
hydrogen became neutral down to $\sim$1.1~R$_*$ with a negligible amount of \ion{H}{II}.
These experiments, therefore, fully supported the notion that ionization from higher levels
is crucial in determining the hydrogen ionization levels in sgB[e] winds.

Despite the fact that hydrogen is ionized in most of our models, our simulations also
prove that predominantly neutral material can exist around sgB[e] stars. 
In model D, for example, hydrogen has already recombined in the equatorial region.
Furthermore, models B and C demonstrated that other species, like helium, can be neutral near 
the stellar surface even if hydrogen is ionized.
It is a realistic expectation, therefore, that conditions suitable for molecule and dust 
formation may be found by more realistic ASTAROTH models.
The task for follow-up studies is to map the parameter space (much more extended than the one
for spherical models) and find their location. 
Our calculations also highlighted the complex nature of sgB[e] envelopes and the limitations of 
simple semi-analytic calculations.
For detailed mapping of the parameter space, self-consistent numerical simulations are necessary. 

Before we discuss our plans for the future, a few words about improving ASTAROTH and the sgB[e] 
models are warranted.  One simplification in our sgB[e] models is the Sobolev approximation
for the line transfer. At low latitudes the flow velocities were close to the ion thermal speed where
the Sobolev approximation may no longer be accurate.
We checked our results against CMFGEN models created from selected latitudes 
of the 2D models (e.g., pole or equator) that were using full line transfer. 
We saw no indication in these tests that departure
from the Sobolev approximation would give a fundamentally different result.
Note, that ASTAROTH and CMFGEN are very different codes and they use different solution 
techniques! 

\begin{figure}
\centering
\includegraphics[angle=270, scale= 0.3]{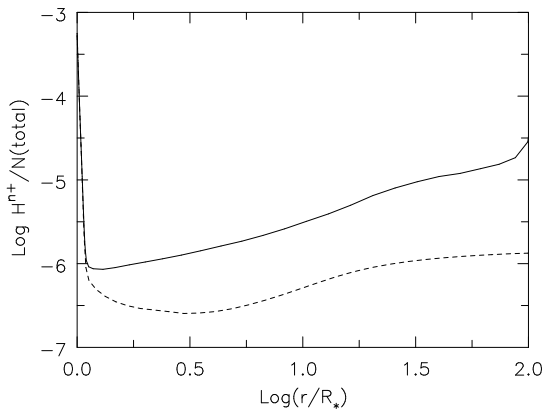}
\caption{
The neutral hydrogen fraction as a function of radius in a spherical CMFGEN 
model created from the equatorial region of model A (dashed line); and in 
the actual equatorial region of our model A (solid line). 
}  \label{fig:comp}
\end{figure}

Fig.~\ref{fig:comp} shows that the neutral hydrogen fraction is low in both the 
spherical non-Sobolev model and in the equator of the actual 2D calculation (model A). 
Such a comparison is not entirely valid, but it can reveal large inconsistencies if present.
The results of CMFGEN and ASTAROTH, on the other hand, were in general agreement.
The ionization level is approximately an order of magnitude lower in 2D since the radiation scatters 
off the dense disk preferentially toward the polar directions.
This comparison also illustrates why modeling 2D problems with spherical codes, latitude by 
latitude, is not sufficient.  Recreating this preferential scattering in a spherical context is impossible.
 
Another simplification is that the effects of rotation were neglected in the simulations presented here.
This is a significant omission if sgB[e] stars are fast rotators, as thought.
\citet{kra06} found important changes when they introduced these effects in their
earlier models \citep{kra03} and we expect the same for our simulations.
Rotation has four effects: First, rotation affects the dynamics of the circumstellar
envelope, and the resultant line profiles. 
Second, rapid rotation causes gravity darkening such that the star is hotter at the 
poles than at the equator \citep{vze24}. 
Third, very rapid rotation causes the star to become oblate. 
Fourth, rotation alters the escape probabilities and 
thus the occupation numbers of various levels.
The inclusion of the fourth and the first two effects into ASTAROTH is easy; the third is difficult.

We also anticipate significant changes when improving the atomic models.
Iron is an especially important element to include because line-blanketing can seriously alter
the radiation field, and hence influence the ionization of hydrogen and helium.
Unfortunately, this is a time consuming task because iron has tens-of-thousands of lines, and
in order to allow for blanketing the line transfer must be done in the CMF. An alternative is to
introduce the effects of line-blanketing in an approximate manner
\citep[e.g., like in FASTWIND,][]{pul05}.

It is also very important to explore various dynamical concepts for sgB[e] envelopes and to
include the non-radial velocity field.
As mentioned in \S\ref{section:intro}, the bi-stability mechanism is only one
explanation for the observed bi-modal structure of the envelope.
An important alternative is the Keplerian disk model which gained some support 
by recent hydrodynamical simulations \citep[e.g.,][]{mad07}. 
The problem with models based on the classical line driven wind theory \citep{CAK}, like the 
bi-stability model, is that accelerating the large mass in the equatorial regions requires multiple 
scattering which is difficult to maintain if the photons can easily escape in the polar directions.
This is the very same effect that allowed lower hydrogen ionization in 2D than in the corresponding 
spherical models (see Fig.~\ref{fig:comp}).
Furthermore, in hydrodynamical simulation of stars that are 
not close to the Eddington limit \citep{owo98}, 
the enhanced flux at the pole of a rapidly rotating star tends to create
a bipolar outflow rather than an equatorial disk.

In follow-up studies, we intend to improve ASTAROTH and will perform spectral analysis of 
specific sgB[e] stars.
First, we plan to introduce non-radial velocities and the effect of rotation.
We will also use a more realistic atomic model.
The inclusion of iron blanketing is not trivial (primarily because of computational effort), so
we will first concentrate on a realistic H, He, C, N, and O models.
With these improvements we will undertake a spectroscopic analysis of \object{R~126}, and will use the 
well-observed sample of LMC/SMC sgB[e] stars \citep{zic85, zic86, zic92, zic96}
to fully explore the allowed parameter ranges of the different sgB[e] models (Keplerian disk, bi-stability, 
etc) and look for observable differences.

\section{Summary}\label{section:sum}

We performed numerical calculations of hydrogen and helium ionization in sgB[e] envelopes
with ASTAROTH, our newly developed 2D stellar atmosphere code,
and explored the effect of mass-loss and T$_{eff}$ by sampling the upper end of the sgB[e] parameter
range.  We also presented theoretical profiles for H$\alpha$ and the \ion{C}{IV} $\lambda\lambda$1550
doublet and examined their behavior for different viewing angles, and as a function of the
adopted stellar and envelope parameters.
Our work yielded the following results:

\begin{enumerate}

\item 
We found that it is much harder to form neutral hydrogen in these sgB[e] envelopes than
suggested by previous studies.  
The reason for the discrepancy is the nebular approximation used by the earlier studies;
our calculation shows that ionization from excited 
states of hydrogen is very important in the envelopes of sgB[e] stars and highlights the need for 
self-consistent numerical simulations to study these objects.

\item Of the four models considered, only one formed a neutral hydrogen disk. Even in this case
the disk did not extend to the stellar surface --- it was truncated at $\sim$2-3~$R_*$.
A neutral hydrogen disk formed  for the combination 
of high mass-loss ($>10^{-5}$~M$_{\odot}$~yr$^{-1}$) and low effective temperature (T$_{eff}<$18,000~K).

\item An equatorial disk predominantly neutral in helium forms for all but the lowest mass-loss
rates ($<10^{-6}$~M$_{\odot}$~yr$^{-1}$).

\item
The polar regions remain highly-ionized (e.g., \ion{C}{IV}) in all models.

\item 
The simulated spectra showed many familiar features of sgB[e] observations, like the narrow double-peak 
H$\alpha$ emission and UV resonance line P~Cygni profiles without emission.
We found that H$\alpha$ forms mainly in the disk, and the strength of the continuum scales with
the projected area of the disk--wind interface.

\end{enumerate} 

In the future we will include the effects of rotation (initially gravity darkening, 
non-radial velocities; later surface distortions) and improve our atomic model by including N, O, and more levels of C. 
We will further explore the parameter range, allowed for sgB[e] stars, by using the well-observed LMC/SMC
sample of these stars. An obvious question to be addressed by these studies is -- what is the
structure of the disk (Keplerian, or dense outflow)?

\begin{acknowledgements}

This research was supported by NSF grant AST-0507328 and by Mexico CONACyT grant 42809 (L.~N. Georgiev).

\end{acknowledgements}

%----------------------------------------------------------------%
% Bibliography                                                         %
%----------------------------------------------------------------%

%\bibliographystyle{aa}
%\bibliography{8293}

\end{document}